\documentclass[twocolumn]{aastex62}

\usepackage{xcolor}
\usepackage{subfigure}
\usepackage{appendix}
\usepackage{threeparttable}
\usepackage{multirow}

\graphicspath{{./}{figures/}}

\begin{document}

\title{Multiwavelength Afterglow Analysis of GRB\,221009A:\\Unveiling the Evolution of a Cooling Break in a Wind-like Medium} 
\correspondingauthor{Donggeun Tak (\href{donggeun.tak@gmail.com}{donggeun.tak@gmail.com})}

\author[0000-0002-9852-2469]{Donggeun Tak}
\affiliation{SNU Astronomy Research Center, Seoul National University, Seoul 08826, Republic of Korea}
\affiliation{Korea Astronomy and Space Science Institute, Daejeon 34055, Republic of Korea}

\author{Z. Lucas Uhm}
\affiliation{Korea Astronomy and Space Science Institute, Daejeon 34055, Republic of Korea}

\author[0000-0002-6639-6533]{Gregory S. H. Paek}
\affiliation{SNU Astronomy Research Center, Seoul National University, Seoul 08826, Republic of Korea}
\affiliation{Astronomy Program, Department of Physics and Astronomy, Seoul National University, Gwanak-gu, Seoul 08826, Republic of Korea}

\author[0000-0002-8537-6714]{Myungshin Im}
\affiliation{SNU Astronomy Research Center, Seoul National University, Seoul 08826, Republic of Korea}
\affiliation{Astronomy Program, Department of Physics and Astronomy, Seoul National University, Gwanak-gu, Seoul 08826, Republic of Korea}

\author[0000-0003-1250-7872]{Makoto Arimoto}
\affiliation{Faculty of Mathematics and Physics, Institute of Science and Engineering, Kanazawa University, Kakuma,
Kanazawa, Ishikawa 920-1192, Japan}

\author[0000-0003-4422-6426]{Hyeonho Choi}
\affiliation{SNU Astronomy Research Center, Seoul National University, Seoul 08826, Republic of Korea}
\affiliation{Astronomy Program, Department of Physics and Astronomy, Seoul National University, Gwanak-gu, Seoul 08826, Republic of Korea}

\author[0000-0002-0070-1582]{Sophia Kim}
\affiliation{SNU Astronomy Research Center, Seoul National University, Seoul 08826, Republic of Korea}
\affiliation{Astronomy Program, Department of Physics and Astronomy, Seoul National University, Gwanak-gu, Seoul 08826, Republic of Korea}

\author[0000-0002-5448-7577]{Nicola Omodei}
\affiliation{W. W. Hansen Experimental Physics Laboratory, Kavli Institute for Particle Astrophysics and Cosmology, Department of Physics and SLAC National Accelerator Laboratory, Stanford University, Stanford, CA 94305, USA}

\author[0000-0002-4744-9898]{Judith Racusin}
\affiliation{NASA Goddard Space Flight Center, Greenbelt, MD 20771, USA}

\author[0000-0001-7082-6009]{Yuji Urata}
\affiliation{Institute of Astronomy, National Central University, Chung-Li 32054, Taiwan;}

\author[0000-0002-9725-2524]{Bing Zhang}
\affiliation{The Nevada Center for Astrophysics, University of Nevada, Las Vegas, NV 89154, USA}
\affiliation{Department of Physics and Astronomy, University of Nevada, Las Vegas, NV 89154, USA}

\begin{abstract}

Gamma-ray bursts (GRBs) are the most energetic explosions in the universe, and their afterglow emission provides an opportunity to probe the physics of relativistic shock waves in an extreme environment. Several key pieces for completing the picture of the GRB afterglow physics are still missing, including jet properties, emission mechanism, and particle acceleration. Here we present a study of the afterglow emission of GRB\,221009A, the most energetic GRB ever observed. Using optical, X-ray, and gamma-ray data up to approximately two days after the trigger, we trace the evolution of the multi-wavelength spectrum and the physical parameters behind the emission process. The broadband spectrum is consistent with the synchrotron emission emitted by relativistic electrons with its index of $p = 2.29\pm 0.02$. We identify a break energy at keV and an exponential cutoff at GeV in the observed multi-wavelength spectrum. The break energy increases in time from $16.0_{-4.9}^{+7.1}$ keV at 0.65 days to $46.8_{-15.5}^{+25.0}$ keV at 1.68 days, favoring a stellar wind-like profile of the circumburst medium with $k=2.4\pm0.1$ as in $\rho (r) \propto r^{-k}$. The high-energy attenuation at around 0.4 to 4 GeV is attributed to the maximum of the particle acceleration in the relativistic shock wave. This study confirms that the synchrotron process can explain the multi-wavelength afterglow emission and its evolution. 
\end{abstract}


\section{Introduction}

Gamma-ray bursts (GRBs) are the most energetic electromagnetic explosions occurring in the universe, whose afterglow emission is powered by a relativistic blast wave sweeping up a surrounding ambient medium \citep{meszarosrees97,sari98}. The broadband afterglow emission stems from synchrotron radiation of non-thermal electrons accelerated at the shock fronts of the blast wave \citep{rybicki79}. These accelerated electrons are assumed to form a power-law spectrum with an index $p$ above a minimum Lorentz factor $\gamma_m$ and then cool down to a Lorentz factor $\gamma_c$ within the dynamical timescale of the relativistic blast wave. Depending on the synchrotron cooling rate, the relative location of two synchrotron frequencies $\nu_m$ and $\nu_c$, which correspond to $\gamma_m$ and $\gamma_c$, respectively, present two different cooling regimes for the broadband synchrotron spectra \citep{sari98}: the slow-cooling regime ($\nu_m < \nu_c$) and the fast-cooling regime ($\nu_c < \nu_m$).

This widely-used model for GRB afterglows assumes a single zone for the entire shocked-medium endowed with the same energy density and magnetic field strength and predicts a sharp spectral break at $\nu_m$ and $\nu_c$. \cite{granotsari02} relaxed this assumption and showed that these spectral breaks should be very smooth, and \cite{uhm14b} identified the origin of smoothing as coming from different cooling histories of co-moving magnetic field strength in different parts of the shocked material. As the blast wave propagates through the ambient medium, the frequency of spectral breaks evolves distinctly in time: $\nu_m \propto t^{-3/2}$ for both the interstellar medium (ISM; $\rho \propto$ const) and the wind medium ($\rho \propto r^{-2}$), and $\nu_c \propto t^{-1/2}$ for ISM and $\nu_c \propto t^{1/2}$ for wind \citep{sari98,granotsari02}. Here, the time $t$ is measured in the observer frame.

The spectral breaks in the optical to X-ray energy range are rarely seen in observations of GRB afterglows. One plausible reason for this is probably due to the very smooth nature of spectral breaks occurring over few orders of magnitude in frequency \citep{granotsari02,uhm14b}, while the observations do not usually cover such a wide range of afterglow spectra without sparseness of data. Even though there have been some hints or suggestions on the possible detection of a cooling break $\nu_c$ \citep[e.g.,][]{filgas11,kouveliotou13,ajello20}, the measured evolution of their spectral break was barely consistent with expectations of the cooling break in the ISM or the wind medium scenario, requiring an additional rationale.

GRB\,221009A exhibits exceptional luminosity with a clear TeV afterglow characterized by the detection of over 64,000 photons above 0.2 TeV, as observed by the Large High Altitude Air Shower Observatory \citep{LHAASO2023}. The TeV afterglow suggests a two-component jet picture with a narrow core consistent with being Poynting-flux-dominated, while a wide structured wing is matter-dominated \citep[e.g.,][]{Zhang2024, Zheng2024}. Even though the early TeV afterglow suggests a uniform density, subsequent afterglow analyses indicate the necessity of a wind-like medium at large radii \citep[e.g.,][]{Ren2023,Zheng2024}. Due to the extraordinary interest in this event, extensive follow-up observations were conducted, enabling a comprehensive broadband analysis \citep[e.g.,][]{Laskar2023, OConnor2023, Ren2024, Klinger2024}.

In this paper, we perform a multi-wavelength analysis on the afterglow observations of GRB 221009A, spanning the optical to GeV energy band, and identify a cooling break $\nu_c$ in the broadband spectra. Also, the temporal evolution of the cooling break is traced agreeing with the afterglow model of the wind-like medium. Throughout this paper, the spectral flux $F_\nu$ depends on time and frequency as $F_\nu \propto t^{-\alpha}\nu^{-\beta}$, where $\alpha$ and $\beta$ are temporal and spectral indices, respectively. We adopt $z=0.151$ for the redshift of GRB\,221009A, reported by X-shooter of the Very Large Telescope \citep{GCN32648}.

\section{Observations}\label{sec:obs}
\begin{figure*}[t]
  \centering
  \includegraphics[scale=0.75]{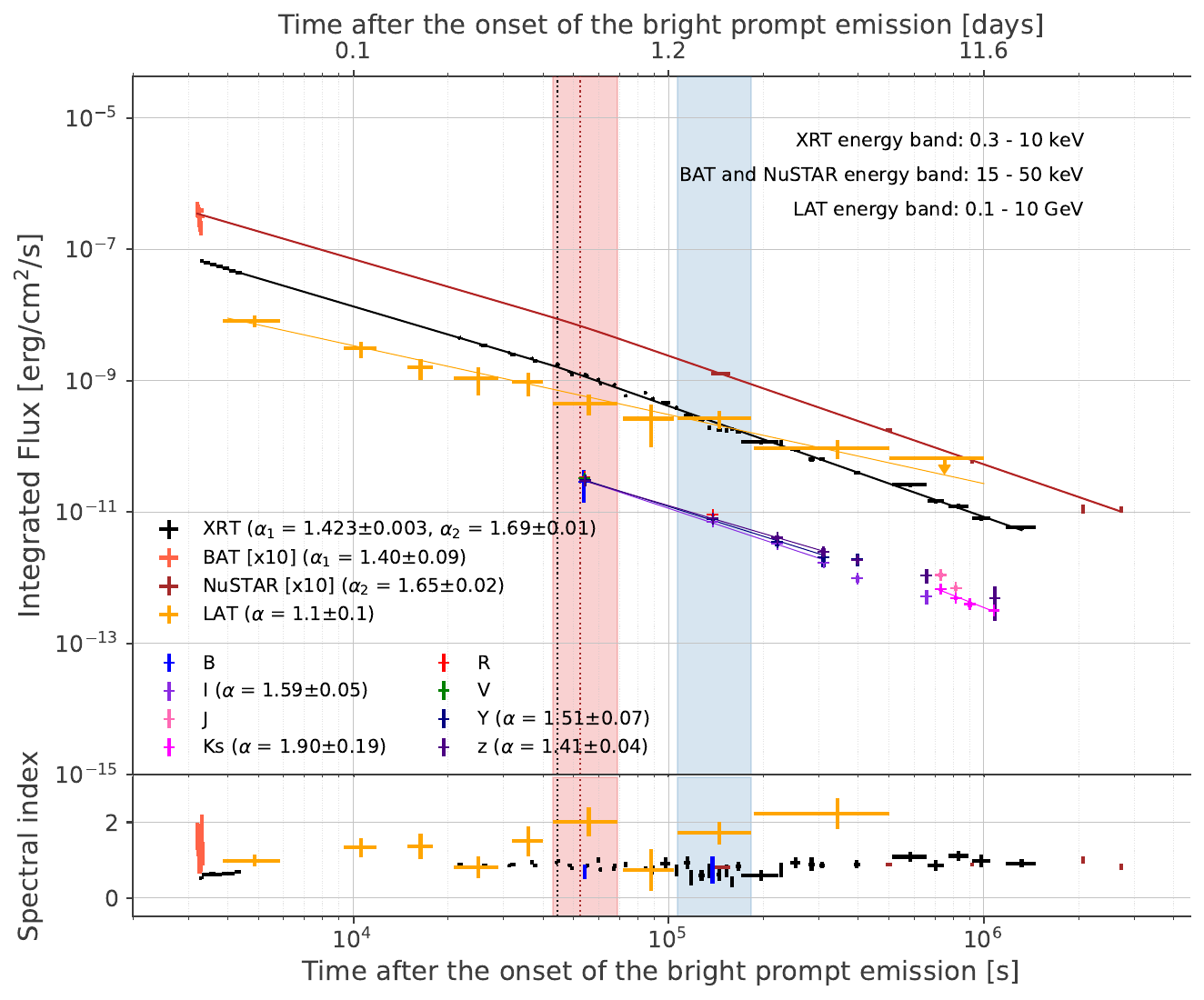}
  \caption{Multi-wavelength afterglow lightcurves and the photon index evolutions for the optical (7 bands), \textit{Swift}-XRT (0.5--10 keV; black), \textit{Swift}-BAT and NuSTAR (15--50 keV; light brown and brown, respectively), and \textit{Fermi}-LAT (0.1--10 GeV; orange) data from GRB\,221009A. The flux for each instrument is calculated from the best-fit model for each time interval in the spectral analysis with a proper method as described in Section~\ref{sec:obs}. To remove the $t_0$ effect, we shift all lightcurves by 150 s in consideration of the onset of the bright main emission of GRB\,221009A. The light curve for the \textit{Fermi}-LAT extended emission is well described with a power law with its index of $1.1\pm0.1$. The \textit{Swift}-XRT and \textit{Swift}-BAT/NuSTAR lightcurves show a consistent, smooth transition in its temporal index from about 1.4 to 1.7, where their smooth breaks emerge at $4.4\times10^5$ s (0.5 days; black dotted line) and $5.2\times10^5$ s (0.6 days; brown dotted line), respectively. The optical/NIR lightcurves decay in time with the power law. The two shaded regions (red and blue) indicate the time intervals where we perform the dedicated multi-wavelength analysis: 0.5--0.8 days and 1.2--2.1 days. The optical photon indices calculated for the two time intervals are depicted in blue. }
  \label{fig:lc}
\end{figure*}
\subsection{Ground-based Optical/NIR follow-up observations}

We observe the GRB\,221009A afterglow with the optical/NIR telescopes in the Small Telescope Network of Korea \citep[SomangNet;][]{2021JKAS...54...89I}, starting from 15 hours after the \textit{Fermi}-GBM trigger (2022 October 09 at 13:16:59.99 UT; $t_{0, \rm GBM}$) for 13 days. SomangNet consists of 1-m class telescopes all around the world, including the Lemmonsan Optical Astronomy Observatory (LOAO; 1.0-m), Bohyunsan Optical Astronomy Observatory (BOAO; 1.8-m), SNU RASA36 (RASA36; 0.36-m), SNU Astronomical Observatory (SAO; 1.0-m), KIAS Chamnun Telescope (KCT; 0.36-m), and Lee Sang Gak Telescope (LSGT; 0.43-m). Detailed information about these telescopes can be found in \cite{2021JKAS...54...89I}. 

Optical/NIR images from the LOAO ($B, V, R, I,z,$ and $Y$ filters), RASA36 ($r$ filter), and SAO ($R$ and $I$ filters) are reduced by an automatic image process pipeline, {\tt gpPy} \citep[Paek et al. in prep;][]{gppy}. Figure~\ref{fig:phot} in Appendix~\ref{sec:opt} shows the location of the observed afterglow. We perform calibration using the point sources from the Pan-STARRS1 $\rm 3\pi$ survey DR1 catalog \citep[PS1; ][]{2016arXiv161205560C} located at a distance of 75\% of the image size from the GRB afterglow within magnitudes ranging from 12 to 20. The NIR images from BOAO ($J\: \&\:Ks$) are reduced and stacked with an {\tt xdimsum} task in {\tt IRAF} \citep{1986SPIE..627..733T}. Photometry calibration is done with point sources in the 2MASS All-Sky Catalog of Point Sources \citep{2MASS}. 

Before performing the photometry, to eliminate light from a nearby galaxy, we subtract reference images from the observed images using the {\tt HOTPANTS} software \citep{2015ascl.soft04004B}, except for two LOAO 1.0-m images ($z$ and $Y$ bands) and two BOAO 1.8-m images (J and Ks bands) due to absence of reference images. For the reference images, we use the late-time images when the signal from this GRB is negligible. If such images are not available, a PS1 image, taken at a wavelength similar to the observed image, is used \citep{2016arXiv161205560C}. We measure aperture photometry with the {\tt SourceEXtractor} \citep{1996A&AS..117..393B}. Table~\ref{tab:phot} in Appendix~\ref{sec:opt} provides a comprehensive summary of the optical and NIR observations newly released in this work before extinction corrections. We correct for the Galactic extinction, following \cite{2011ApJ...737..103S}.  For the host galaxy extinction, we adopt the Milky Way (MW) extinction law \citep[$\eta$;][]{Pei1992}\footnote{The difference in the extinction laws between MW, Large Magellanic Cloud, and Small Magellanic Cloud is insignificant in our optical/NIR energy bands.} and follow the model by \cite{Kann2006}, $F_\nu = F_{\nu, 0}\exp(-A_\nu\eta(\nu_{\rm rest})/1.086)$, where the correction factor $A_\nu$ is unknown and varies one to another GRB. 

\subsection{\textit{Swift} observations}
GRB\,221009A (initially named Swift J1913.1+1946) triggered the Burst Alert Telescope on board the \textit{Neil Gehrels Swift Observatory} \citep[\textit{Swift}-BAT; 15--150 keV;][]{Barthelmy05} at 2022-10-09 14:10:17.995 UT \citep[$t_{0, \rm BAT}$ = $t_{0, \rm GBM}$ + 53.3 min;][]{GCN32632}. The X-Ray Telescope \citep[\textit{Swift}-XRT; 0.3--10 keV;][]{Burrows05a} started to observe the burst about 143 s after $t_{0, \rm BAT}$, and the UltraViolet Optical Telescope \citep[\textit{Swift}-UVOT; 170-650 nm;][]{Roming05} also began its observations of GRB\,221009A from $t_{0, \rm BAT}$ + 179 s. Both \textit{Swift}-XRT and \textit{Swift}-UVOT detected the X-ray and optical counterparts of the burst, respectively, and localized it at (R.A., Dec.) = (288.26452, 19.77350) degrees with a 90\%-confidence error radius of about 0.61 arcsec. Continued observations from $t_{0, \rm BAT}$ + 0.99 days to $t_{0, \rm BAT}$+1.46 days discovered multiple expanding dust rings in the energy band from 0.3 to 10 keV, centered at the GRB position, which are extended from about 2.5 to 7.5 arcmin \citep{GCN32680}. For further details on the Swift observation of GRB 221009A, see \cite{Williams2023}.

Both fluxes and indices of \textit{Swift}-BAT (15--50 keV) and \textit{Swift}-XRT (0.3--10 keV) are acquired from the UK Swift Science Data Centre (Figure~\ref{fig:lc}). For the multi-wavelength spectral analysis (Section~\ref{sec:analysis}), however, in order to avoid possible absorption effects by the multiple dust rings, we manually select background regions with {\tt ds9} far outside the dust-ring region and reduce the dataset with {\tt HEASoft} (v6.31.1)\footnote{{\tt xselect, xrtmkarf}, and {\tt xrtexpomap}}. A \textit{Swift}-XRT spectral analysis is performed with {\tt xspec} (v12.13.0c), where its spectrum from 0.3 to 10 keV is modeled with the power-law function taking into account Galactic \citep[$N_{\rm H} = 5.38 \times 10^{21}$ atoms$/cm^{2}$;][]{Willingale2013}\footnote{The value of $N_{\rm H}$ refers to the Galactic hydrogen absorbing column in a given direction.} and intrinsic absorptions at the GRB location. 

\subsection{NuSTAR observations}
The Nuclear Spectroscopic Telescope Array \citep[NuSTAR; 3--79 keV;][]{Harrison2013} conducted follow-up observations on GRB\,221009A, staring from 2022-10-11 03:10:07 UTC \citep[$t_{0, \rm GBM}$ + 38 hours;][]{GCN32695}. Five NuSTAR observations have been made, starting approximately 1.58, 5.67, 10.5, 23.7, and 31.4 days after $t_{0, \rm GBM}$, for a duration of 17.4, 22.5, 20.4, 21.3, and 40.8 ks (live times), respectively\footnote{Note that during the first observation epoch, M-size solar flares occurred so that we filter out the corresponding time intervals.}.  

To reduce the NuSTAR data, we use a module {\tt nuproducts}, implemented in {\tt HEASoft}, to extract the high-level data after defining the good time interval (GTI) with {\tt xselect} and proper regions for the burst and backgrounds with {\tt ds9}. Among the two telescopes of NuSTAR, we take ``FPMA" data; note that our result remains consistent independent of the selection of telescopes. The NuSTAR spectrum is also fitted with the power-law model, considering the same level of the galactic absorption used in the \textit{Swift}-XRT analysis.

\subsection{\textit{Fermi}-LAT observations}
The Gamma-ray Burst Monitor on board the \textit{Fermi Gamma-ray Space Telescope} \citep[\textit{Fermi}-GBM; 8 keV--40 MeV;][]{Meegan2009} was triggered by a short precursor of GRB\,221009A \citep{GCN32642}. The extraordinarily bright prompt emission of GRB\,221009A was observed about 180 s after $t_{0, \rm GBM}$, which lasts more than 300 s ($t_{90} = 327 s$). The \textit{Fermi} Large Area Telescope \citep[\textit{Fermi}-LAT; 100 MeV--300 GeV;][]{Atwood09} detected the high-energy counterpart of GRB\,221009A \citep{GCN32637}, where the high-energy emission continued for about 6 days (Figure~\ref{fig:lc}). Since GRB\,221009A is extremely bright at its peak, \textit{Fermi}-GBM and \textit{Fermi}-LAT suffered pile-up and count-rate saturation, resulting in the bad time interval (BTI) in both instruments, e.g., the BTI of the \textit{Fermi}-LAT is from $t_{0, \rm GBM}$ + 203 s to $t_{0, \rm GBM}$ + 294 s \citep{Axelsson2024}. 

We analyze the extended \textit{Fermi}-LAT emission, starting from 4000 s after $t_{0, \rm GBM}$, with a {\tt python} package, {\tt fermipy} (v1.2). Since this burst is unusually bright, we select \textit{Source} class events of the  Pass 8 data (P8R3) and its corresponding instrument response functions (P8R3\_SOURCE\_V3). We select events from the region of interest (ROI) of 12x12 degree and apply the zenith cut of 100 degree. GRB\,221009A is modeled with a power-law function, and its background includes contributions from the \textit{Fermi}-LAT catalog sources (4FGL-DR3) within the ROI, the isotropic (iso\_P8R3\_SOURCE\_V3\_v1.txt) and galactic (gll\_iem\_v07.fits) diffuse backgrounds \citep{4FGL}. When a time interval is short ($<$ 3000 s), all nuisance parameters in the background components are fixed, except for the normalization of the isotropic diffuse background. Otherwise, we additionally fit the normalization of nearby sources (i.e., distance from the position of the burst is less than 3.0 degrees) or sources with test statistic (TS) higher than 5. For a more detailed analysis of the Fermi-LAT observation, see \cite{Axelsson2024}.

\section{Broadband afterglow}\label{sec:result}

\begin{figure*}[t]
  \centering
  \subfigure{\includegraphics[width=0.35\paperwidth]{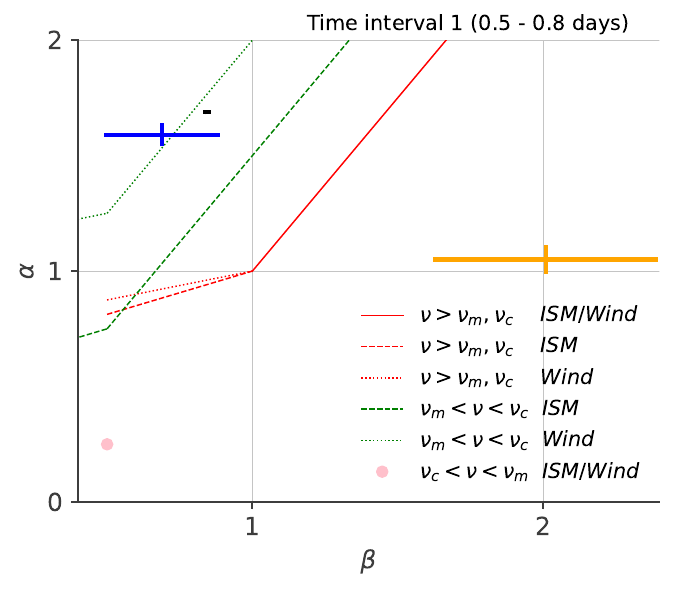}}  \subfigure{\includegraphics[width=0.35\paperwidth]{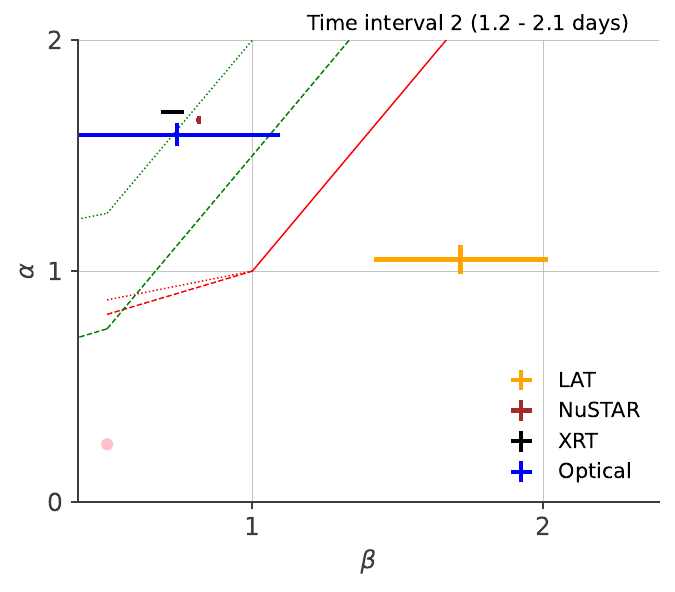}}
  \caption{Test of closure relations with observed temporal and spectral indices from the optical, \textit{Swift}-XRT, NuSTAR, and \textit{Fermi}-LAT data. The left and right panels are the results from the first (0.5--0.8 days) and second (1.2--2.1 days) time intervals, respectively. The data points with 68\% errors are plotted on top of the closure relations: optical (blue), Swift-XRT (black), NuSTAR (brown), and \textit{Fermi}-LAT (orange). The color of the closure relations differentiates a cooling regime; red is used for the fast cooling ($\nu > \nu_m, \nu_c$), green for the slow cooling ($\nu_m < \nu < \nu_c$), and orange for $\nu_c < \nu < \nu_m$. The line style is related to the surrounding environment property; a solid line is used for an undetermined environment, a dashed line for ISM, and a dotted line for wind. }
  \label{fig:cr}
\end{figure*}

\begin{table*}
\small
\centering 
	\begin{tabular}{c c | c c c | c c c}
    \hline\hline
Instrument &  Energy band & Epoch & $\alpha$ & $\beta$ & cooling regime & $k$ & $p$ \\\hline\hline 
\multirow{2}{*}{LOAO} & \multirow{2}{*}{1.3--2.9 eV} & 1 & \multirow{2}{*}{$1.59 \pm 0.05$}\footnote{We take the temporal index obtained from the \textit{I} band that is the middle of the six LOAO bands.} & $0.69 \pm 0.20$\footnote{We assume $A_\nu$ of 0.15 based on \cite{Li2018}.} & \multirow{2}{*}{$\nu_m<\nu<\nu_c$} & $2.10 \pm 0.54$ & $2.37 \pm 0.40$ \\
 & & 2 & & $0.73 \pm 0.36$\textsuperscript{b} & & $1.97 \pm 1.11$ & $2.48 \pm 0.72$ \\\hline
\multirow{2}{*}{\textit{Swift}-XRT} & \multirow{2}{*}{0.3--10 keV} & 1 & \multirow{2}{*}{$1.688\pm0.010$} & $0.843\pm0.014$ & \multirow{2}{*}{$\nu_m<\nu<\nu_c$} & $1.833\pm0.054$ & $2.687\pm0.028$ \\
 & & 2 & & $0.723\pm0.040$ & & $2.187\pm0.099$ & $2.447\pm0.079$ \\\hline
NuSTAR & 3--79 keV & 2 & $1.654 \pm 0.018$ & $0.815 \pm 0.009$ & $\nu_m<\nu<\nu_c$ & $1.853 \pm 0.052$ & $2.630 \pm 0.018$ 
\\\hline
\multirow{2}{*}{\textit{Fermi}-LAT} & \multirow{2}{*}{0.1--10 GeV} & 1 & \multirow{2}{*}{$1.04 \pm 0.06$} & $2.01\pm 0.39$ & \multirow{2}{*}{$\nu>\nu_m, \nu_c$} & - & $4.01 \pm 0.77$ \\
& & 2 & & $1.72 \pm 0.30$ & & - & $3.43 \pm 0.60$\\\hline\hline
\multicolumn{5}{l}{Errors correspond to 1-$\sigma$ confidence region.} & Weighted average & $1.89\pm0.15$ & $2.64\pm0.08$ \\
\end{tabular}
    \caption{Analysis results of individual observations. Columns 1 to 2 show each instrument and its energy band. The next three columns give the temporal ($\alpha$) and spectral ($\beta$) indices from the two time intervals; epoch 1 is 0.5--0.8 days, and epoch 2 is 1.2--2.1 days. The final three columns show the microphysical parameters, density profile index ($k$), and electron spectral index ($p$) obtained from testing the EFS model \citep{sari98}.}
    \label{tab:individual}
\end{table*}

The main emission phase of GRB\,221009A starts about 150 seconds after $t_{0, \rm GBM}$ that corresponds to a weak precursor of the burst. Since a temporal index is highly dependent on the selection of the trigger time \citep[the so-called $t_0$ effect;][]{zhang06}, we need to take the $t_0$ effect into account to get a proper temporal index. Assuming that the main emission phase is accompanied with the onset of the afterglow emission of GRB\,221009A, we shift the lightcurve by 150 s, $t_{0, \rm onset} = t_{0, \rm GBM} + 150 s$ and compute the temporal index. We emphasize that this effect may significantly influence the temporal fit in this analysis, particularly affecting the XRT data fitting.

Figure~\ref{fig:lc} shows the multi-wavelength afterglow lightcurve, starting from 3200 s after $t_{0, \rm GBM}$, with the evolution of the photon index measured from each energy band. In the upper panel, the optical/NIR lightcurves from various energy bands are characterized by the power law with indices of $1.59\pm0.05$ (\textit{I} band), $1.51\pm0.07$ (\textit{Y} band), $1.41\pm0.04$ (\textit{z} band), and $1.90\pm0.19$ (\textit{Ks} band). While the temporal indices from the \textit{I}, \textit{Y}, and \textit{z} bands are consistent, the late-time temporal index obtained from the \textit{Ks} band differs, possibly due to the supernova counterpart of GRB\,221009A \citep{GCN.32769}. Note that this is unlikely to originate from a jet break \citep{Laskar2023}, as the achromatic change in the temporal slope is not observed in other wavelengths. In the bottom panel of Figure~\ref{fig:lc}, the high-energy photon index tends to soften in time, whereas those from other energy bands remain constant. 


The \textit{Swift}-XRT light curve in 0.5--10 keV requires a smooth temporal break with $\alpha_1 = 1.423 \pm 0.003$ and $\alpha_2 = 1.69 \pm 0.01$, respectively. The hard X-ray lightcurve in the energy band from 15 to 50 keV from the Swift-BAT and NuSTAR observations is well described by a smoothly-broken power law with early and late temporal indices of $\alpha_1 = 1.40 \pm 0.02$ and $\alpha_2 = 1.65 \pm 0.02$, respectively. Note that the hard X-ray data are very sparse compared to the other wavelength data. As the afterglow phase evolves, the temporal break shifts to higher energies: in the 0.3–10 keV range, the break occurs approximately 12.3 hours (0.5 days) after $t_{0, \rm onset}$, while in the 15–50 keV range, it is observed around 14.5 hours (0.6 days) \footnote{The smoothness parameters vary between 0.02 and 0.1.}. This X-ray break is also reported in previous studies \citep[e.g.,][]{Laskar2023, OConnor2023}.

The extended high-energy emission detected by \textit{Fermi}-LAT exhibits a shallow decay, with a temporal index of $\alpha = 1.1 \pm 0.1$ in the energy band from 100 MeV to 10 GeV. For comparison, \cite{Axelsson2024} reports a temporal index of $\alpha = 1.27 \pm 0.05$ within the 100 MeV to 100 GeV energy band. This deviation may result from differences in data reduction techniques, the choice of temporal model, or the specific energy range analyzed. For the majority of \textit{Fermi}-LAT GRBs, the \textit{Fermi}-LAT emission typically persists for $10^3$ seconds \citep{2FLGC}. Since this GRB is extremely bright, the spectral index evolution is well determined up to $t_{0, \rm GBM} + 10^5$ s. 

As shown in Figure~\ref{fig:lc}, we have a broadband dataset in the two time intervals: 0.5--0.8 (shaded in red) and 1.2--2.1 days (shaded in blue). Table~\ref{tab:individual} shows the temporal and spectral indices for the two time intervals obtained from the individual data analysis. Here, we fix $A_\nu$ to 0.15 based on the systematic study on the GRB host galaxy by \cite{Li2018} because it is hard to constrain both $\beta$ and $A_\nu$ simultaneously with a narrow energy-band data. Note that the spectral index is highly dependent on $A_\nu$. With the obtained indices, we test the external forward shock (EFS) model by \cite{sari98}. This model predicts a specific relation between the temporal and spectral indices, the so-called closure relation, which is established by a combination of the cooling regime, electron spectral index $p$, and circumburst density profile $\rho(r)$. Following the method in \cite{tak2019}, we test the set of closure relations. As shown in Figure~\ref{fig:cr}, the results from optical, \textit{Swift}-XRT, and NuSTAR observations all converge at a similar point in both time intervals and imply that the optical to X-ray emission can originate from the same emission component; i.e., synchrotron emission in the slow cooling regime ($\nu_m < \nu < \nu_c$) from EFS developed in the wind environment ($k=2$ in $\rho \propto r^{-k}$). In this scenario, we can estimate the exponent of the density profile \citep{Eerten2009}, 
\begin{equation}
    k = \frac{4}{1+\frac{1}{2\alpha-3\beta}}.
\end{equation}
The weighted average of the estimated $k$ is $1.89\pm0.15$, implying the wind-like environment (Table~\ref{tab:individual}). 

Contrarily, the \textit{Fermi}-LAT data from the two time intervals deviate from the others, where their spectral indices are too soft to belong to any of closure relations, suggesting other physical conditions (see Section~\ref{sec:disc} and Appendix~\ref{sec:eb_evol}). However, at 3--4 $\times 10^3$ s, the set of the observed spectral ($\beta \sim 1$) and temporal indices ($\alpha \sim 1$) agrees with the closure relation of $\alpha = (3\beta-1)/2$, which is from the cooling condition of $\nu > \nu_m, \nu_c$ (red solid line). This implies that the broadband emission (possibly from the optical to GeV energy band) can be explained with a simple afterglow picture at the beginning, but the GeV emission deviates at the late phase as the afterglow phase proceeds.

Assuming that the cooling break is well above the X-ray energy band but below the $\gamma$-ray energy band, we convert $\beta$ to $p$ by using $p = 2\beta+1$ (optical, \textit{Swift}-XRT, and NuSTAR) and $p = 2\beta$ (\textit{Fermi}-LAT), resulting in the weighted average of $p = 2.64 \pm 0.08$ (Table~\ref{tab:individual}).

\section{Multi-wavelength spectral analysis}\label{sec:analysis}

\begin{figure*}[t]
  \centering
  \includegraphics[scale=0.7]{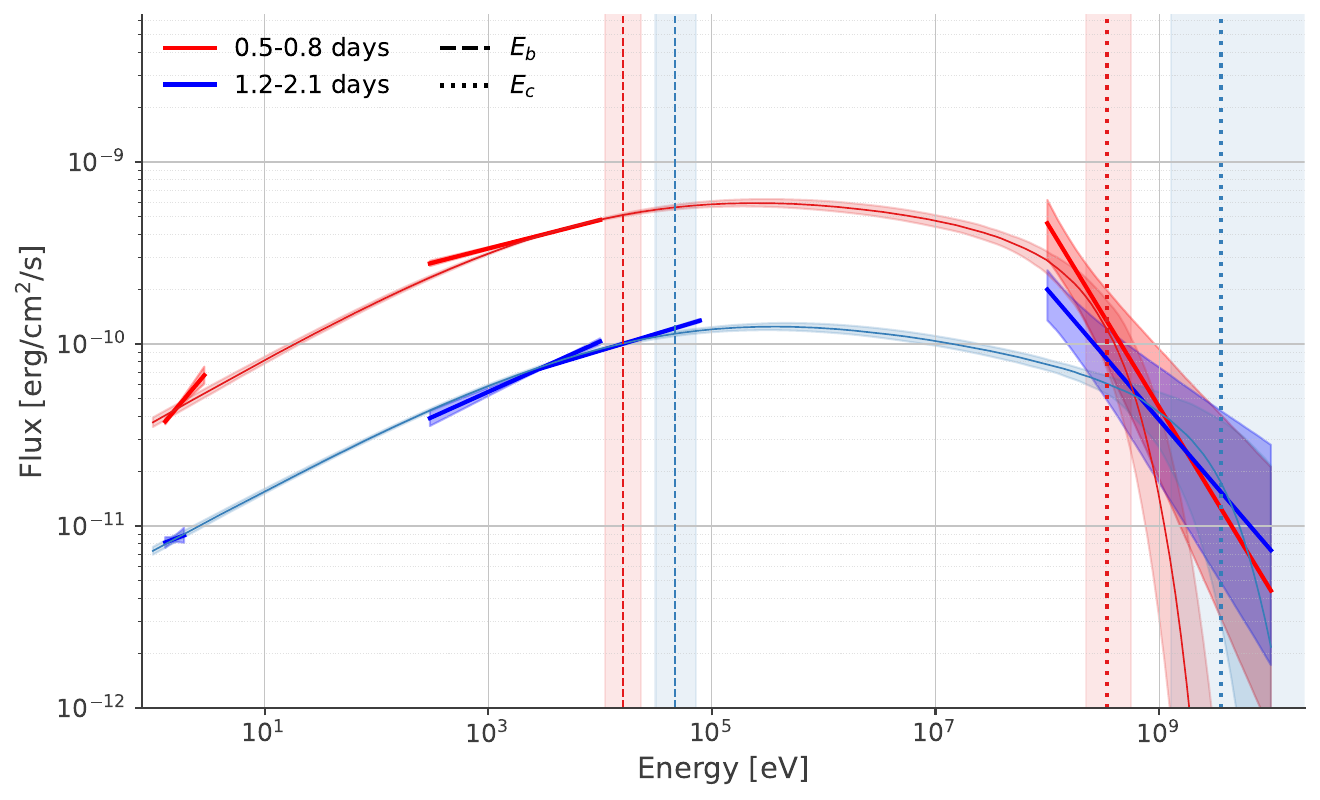}
  \caption{Multi-wavelength SED from optical to $\gamma$-ray energies for two time intervals: 0.5--0.8 days (red) and 1.2--2.1 days (blue). The broadband lines and their contours correspond to the median and 68\% containment of the afterglow model \citep{granotsari02} with an exponential cutoff (Equation~\ref{eq:f}). Likewise, the lines and their contours with deeper colors represent the best-fit power-law model to the data from each individual observation. The vertical dashed and dotted lines and shaded regions denote median and 68\% containment of the break energy ($E_b$) and the cutoff energy ($E_c$), respectively (Table~\ref{tab:sed}).}
  \label{fig:sed}
\end{figure*}

We perform the multi-wavelength analysis for the two time intervals (0.5--0.8 and 1.2--2.1 days) using multiple datasets from the optical, X-ray, and $\gamma$-ray energy bands; in the first time interval, we configure the dataset with LOAO, Swift-XRT, and Fermi-LAT data, and in the second time interval, NuSTAR data is additionally included. For the likelihood analysis, we utilize a {\tt python} package, the Multi-Mission Maximum Likelihood framework \citep[{\tt 3ML}; ][]{3ML_ref}, which is developed for combining and analyzing multiple dataset by exploiting appropriate plugins for each instrument. For the optical data, we develop a plugin compatible with {\tt 3ML}, where the likelihood for the optical data is calculated with $\chi^2$. For the other datasets, we adopt a proper plugin provided by {\tt 3ML}: i.e., OGIPLike for \textit{Swift}-XRT and NuSTAR, and FermipyLike for \textit{Fermi}-LAT. Note that we introduce a constant factor $C_{\rm x}$ to account for the potential of relative calibration uncertainties between \textit{Swift}-XRT and NuSTAR (i.e., normalization).

For the multi-wavelength dataset, we test the simplified afterglow model suggested by \cite{granotsari02} of which the synchrotron emission spectrum from the EFS model is approximated with a smoothly broken power law as follows,
\begin{equation}\label{eq:efs}
    \frac{dN_{\rm EFS}(E)}{dE} = K \left( \frac{\left(\frac{E}{E_b}\right)^{s\Gamma_1}+\left(\frac{E}{E_b}\right)^{s\Gamma_2}}{\left(\frac{{\rm 1 keV}}{E_b}\right)^{s\Gamma_1}+\left(\frac{{\rm 1  keV}}{E_b}\right)^{s\Gamma_2}}\right)^{-1/s},
\end{equation}
where $K$ is the normalization constant at 1 keV, $E_b$ is a break energy, and $\Gamma_1$ and $\Gamma_2$ are low- and high-energy photon indices, respectively. The smoothness parameter $s$ is a function of $p$, where its functional form depends on the cooling regime and surrounding medium profile. In the slow cooling regime ($\nu_m < \nu_c$) with the wind medium, the smoothness parameter is given by $s = 0.80 - 0.03p$, where $p$ can be estimated from the low-energy photon index; $p = 2\Gamma_1-1$. In the EFS model \citep{sari98}, the low- and high-energy spectral indices differ by 0.5 in the slow cooling regime; i.e., $\Delta \beta = 0.5$. Consequently, the high-energy photon index is constrained to be related to the low-energy photon index by $\Gamma_2 = \Gamma_1 + 0.5$. Notably, even without enforcing this constraint, $\Delta \Gamma \sim 0.5$ can still be obtained, consistent with theoretical expectations; however, in such cases, the break energy, $E_b$ is poorly constrained.

Since the \textit{Fermi}-LAT spectral index is too soft to be explained by the afterglow model (see Section~\ref{sec:result}), we additionally introduce an exponential cutoff for the high-energy regime,
\begin{equation}\label{eq:f}
    \frac{dN(E)}{dE} = \frac{dN_{\rm EFS}(E)}{dE} \exp\left(-\frac{E}{E_{\rm c}}\right),
\end{equation}
where $E_{\rm c}$ is a cutoff energy. We remind that this model is used for fitting the GRB\,221009A energy spectrum in consideration of extinctions, absorptions, and backgrounds (Section~\ref{sec:obs}). All nuisance parameters are fixed, except for $N_{\rm H}$ of the intrinsic absorption in the X-ray energy band. Also, we let $A_\nu$ of the host galaxy extinction in the optical band free because it is possible to constrain it with the multi-wavelength dataset.

Following the method used in \cite{Klinger2023}, we take the Bayesian approach to derive the posterior probability of each parameter, assuming the \textit{prior} probability distribution detailed in Appendix~\ref{sec:app}. For the Bayesian sampling procedure, we take advantage of the nested sampling algorithm, MLFriends, with a {\tt python} module, {\tt UltraNest} \citep{Ultranest_ref}. 

Figure~\ref{fig:sed} shows the spectral energy distribution (SED) of our model over-plotted with the individual power-law fits of each instrument. The red and blue shades indicate the results of the first and second time intervals, respectively. We obtain the median and its 68\% containment by randomly sampling 300 posterior results. For an individual power-law fit, we plot the best-fit model with its 68\% containment. Table~\ref{tab:sed} shows the median and 68\% containment of each parameter obtained from the posterior distribution. We obtain $\Gamma_1 \sim$ 1.65 in both time intervals. The characteristic energies, $E_b$ (dashed line) and $E_c$ (dotted line), increase in time. The median values of the nuisance parameters, $N_{\rm H}$ and $A_\nu$, obtained from the two time intervals are dissimilar and may be attributed to systematic observation effects. The relative correction between \textit{Swift}-XRT and NuSTAR is less than 5\%, $C_x \sim 0.95$. The full corner plots of the parameter posterior probability distributions can be found in Appendix~\ref{sec:app}.

\section{Discussion and Conclusion}\label{sec:disc}

\begin{table*}
\centering 
\begin{tabular}{c | c c c c c c c}
    \hline\hline
Epoch & $K$ & $E_b$ & $\Gamma_1$ & $E_c$ & $N_{\rm H}$ & $A_\nu$ & $C_x$\\
 & [$1/cm^2/s/$keV] & [keV] &  & [GeV] & [$10^{22}$ atoms/$cm^2$] \\\hline
1 & $0.196_{-0.003}^{+0.003}$ &
$16_{-5}^{+7}$ &
$1.64_{-0.01}^{+0.01}$ &
$0.34_{-0.12}^{+0.22}$ &
$1.11_{-0.02}^{+0.02}$ &
$0.62_{-0.12}^{+0.13}$\\
2 & $0.037_{-0.001}^{+0.001}$ &
$47_{-16}^{+25}$ &
$1.66_{-0.01}^{+0.01}$ &
$4_{-2}^{+19}$ &
$1.40_{-0.05}^{+0.05}$ &
$0.10_{-0.06}^{+0.10}$ &
$0.95_{-0.01}^{+0.02}$
\\\hline\hline
\end{tabular}
    \caption{Multi-wavelength spectral analysis results for two time intervals: epoch 1 is the time interval from 0.5 to 0.8 days, and epoch 2 is from 1.2 to 2.1 days. Each parameter is presented with a median and its 68 percent containment. The parameter probability distribution of each parameter can be found in Appendix~\ref{sec:app}.}
    \label{tab:sed}
\end{table*}

As shown in Figure~\ref{fig:sed}, the EFS model with an exponential cutoff (Equation~\ref{eq:f}) describes the observed multi-wavelength dataset sufficiently well, given the consistency between the individual and joint-fit modeling results. Our multi-wavelength modeling result can be interpreted as the synchrotron emission from EFS developed in a wind-like medium with the cooling frequency ($\nu_c$) in the X-ray energy band. From $\Gamma_1$, the electron spectral index can be deduced as $p = 2.29 \pm 0.02$, which agrees with theoretical and numerical simulation results \citep[e.g.,][]{meszaros98, spitkovsky08}. We stress that this $p$ value differs from those computed from the individual spectral indices ($p \simeq 2.6$; Table~\ref{tab:individual}). This is because our broadband modeling used an asymptotic spectral index, whereas the individual indices are not; i.e., the resultant spectral index is not affected by the smoothness feature around the cooling break. This $p$ value is, therefore, more likely to reflect the true electron distribution of GRB\,221009A. 

We identified the intriguing evolution of a cooling break, $\nu_c$. The cooling frequency increases in time from $16_{-5}^{+7}$ keV (0.65 days) to $47_{-16}^{+25}$ keV (1.68 days), implying the wind-like medium; in the uniform density medium, $\nu_c$ decreases in time. The probability of superiority derived from the two posterior distributions of $\nu_c$ yields a p-value of 0.036, which corresponds to approximately 1.8 sigma\footnote{Considering the mean value of one time epoch relative to the other posterior distribution, the difference could range from 2.6 to 2.9 sigma.}. This evolution is also seen in the multi-wavelength lightcurve (Figure~\ref{fig:lc}), where the temporal break appears in the \textit{Swift}-XRT energy band first (0.5 days) and then in the \textit{Swift}-BAT/NuSTAR energy band (0.6 days). We emphasize that the chromatic evolution of the light curve confirms that the temporal breaks are not due to a jet break or the jet structure. We remind that since the cooling break is very smooth and broad, it affects the observing bands gradually and slowly, starting from a long distance in SED. Therefore, we expect a gradual and slight steepening in a light curve. 

The density profile index can be estimated from the $\nu_c$ evolution \citep{Eerten2009},
\begin{equation}
   k = \frac{4}{1+\frac{2}{2y+1}},
\end{equation}
where $y = \log{(\nu_{c,2}/\nu_{c,1})} / \log{(t_2/t_1)}$. The subscription in $\nu_c$ and $t$ indicates a time interval, 1 or 2. With this approach, we obtain $k=2.5^{+0.3}_{-0.4}$. This index is not consistent with the classical wind medium ($k=2$). However, this unusual profile can result from a higher mass-loss rate of the central engine compared to the typical Wolf-Rayet star \citep{vanMarle2008, Starling2008} or from the properties of the structured jet in which the burst energy within the viewing cone does not remain constant. From the flux decay at 1 eV, we can also estimate $k$ \citep{Eerten2009},
\begin{equation}
    k = 4+ \frac{8}{4x+3p-5},
\end{equation}
where $x = \log{(F_{\nu, 2}/F_{\nu, 1})} / \log{(t_2/t_1)}$. With $p$ inferred from $\Gamma_1$, we can get $k=2.4\pm0.1$, which is steeper than the wind scenario. This result is consistent with the previous $k$ estimate from the evolution of the cooling break. Note that this $k$ value differs from that of the closure relation test ($k \simeq 1.89$; Table~\ref{tab:individual}), possibly due to the same reason of the $p$ discrepancy.

Additionally, we could constrain the high-energy cutoff $E_c$ of $0.34_{-0.22}^{+0.12}$ at 0.65 days and $4_{-2}^{+19}$ GeV at 1.68 days. While the high-energy cutoff is commonly observed during the prompt emission phase, GRB\,221009A shows its presence in the afterglow approximately 1--2 days for the first time. The exceptional brightness of GRB\,221009A facilitated this discovery. A possible origin of this cutoff is the maximum synchrotron limit due to the inefficient electron acceleration at a high-energy regime. In theory, the maximum synchrotron energy $E_{\rm max}$ is a function of the bulk Lorentz factor $\Gamma_{\rm bulk}$, $E_{\rm max} \sim 100 \, \Gamma_{\rm bulk}/(1+z)$ MeV \citep{Ackermann2014}, which decreases in time, $E_{\rm max} \propto \Gamma_{\rm bulk} \propto t^{-1/4}$ in the wind medium. The observed $E_c$, however, seems to increase in time. This discrepancy can be attributed to a small Poissonian fluctuation in observed high-energy data, but other possibilities that cause the increase of the cutoff energy, such as an unknown physical process or multiple jets, cannot be ruled out. In addition, the photon index in the Fermi-LAT energy band (orange in Figure~\ref{fig:lc}) seems to be softer in time, starting from $t_{0, \rm GBM} + 10^{4}$ s, which can result from the decrease of $E_c$ in time (see Appendix~\ref{sec:eb_evol}). As time goes on, we expect that this exponential cutoff can progressively affect the spectral slope in the low energy band in the same manner. Assuming that the observed $E_c$ corresponds to $E_{\rm max}$, we can approximately estimate $\Gamma_{\rm bulk}$ to be between 4 to 40, around 1--2 days after the trigger. By retracing to the early phase ($\Gamma \propto t^{-1/4}$), we can obtain $\Gamma_{\rm bulk}$ of 250 at $t_{0, \rm GBM} + 100$ s (see Appendix~\ref{sec:eb_evol}).

Although an unusual jet structure or composition for this burst has been suggested \citep[e.g.,][]{LHAASO2023, OConnor2023, Zhang2024}, our analysis indicates that the evolution of GRB\,221009A over the short time scale from 0.5 to 2.1 days is fully consistent with the standard GRB afterglow model, without the need for invoking a complex jet structure. While \cite{OConnor2023} suggested a decreasing cooling break based on the softening of the X-ray spectral index over a longer time interval (1 hour to 40 days), detailed analyses have reported a distinct transition in the burst environment from a uniform to a wind-like medium at early times \citep{Ren2023, Zheng2024}. This indicates a non-monotonic evolution of the cooling break, where it may decrease during the early phase and increase at later times.

Even though the synchrotron self-Compton (SSC) component, which is present at earlier times (about 1--2 hours after $t_{0, \rm GBM}$) as reported by \cite{LHAASO2023}, is not necessary to explain the broadband emission up to 10 GeV in about  1--2 days after $t_{0, \rm GBM}$, we cannot ignore the possibility that this component could be either sub-dominant or active above 10 GeV. Our estimates on the bulk Lorentz factor from the cutoff energy are comparable to those obtained from the deceleration time: e.g., $\Gamma_{\rm bulk} \gtrsim 282$ \citep{Lesage2023} and $\Gamma_{\rm bulk} = 250 \pm 10$ \citep{Axelsson2024}. 

From the multi-wavelength analysis, we obtain the self-consistent result, where the broadband spectrum and its evolution are adequately explained with the synchrotron emission emitted from EFS developed in the wind-like environment. The test of closure relations implies that the cooling break is likely to be located around the X-ray energy band. The broadband joint-fit analysis provides hints on physical parameters and their evolution. This result provides a complete snapshot of the early afterglow emission (after 4000 s) of GRB\,221009A.

\section*{Acknowledgement}
This work was supported by the National Research Foundation of Korea (NRF) grant, No. 2021M3F7A1084525, funded by the Korea government (MSIT). DT is supported by the NRF grant, RS-2024-00343729. DT, MI and GSHP acknowledge the support from the NRF grant, No. 2020R1A2C3011091, and the Korea Astronomy and Space Science Institute (KASI) R\&D program (Project No. 2022-1-860-03) supervised by the Ministry of Science and ICT. MA acknowledges financial support by JSPS KAKENHI Grant No. JP23H04898
and the CHOZEN Project of Kanazawa University. The \textit{Fermi}-LAT Collaboration acknowledges support for LAT development, operation and data analysis from NASA and DOE (United States), CEA/Irfu and IN2P3/CNRS (France), ASI and INFN (Italy), MEXT, KEK, and JAXA (Japan), and the K.A.~Wallenberg Foundation, the Swedish Research Council and the National Space Board (Sweden). Science analysis support in the operations phase from INAF (Italy) and CNES (France) is also gratefully acknowledged. This work was performed in part under DOE Contract DE-AC02-76SF00515. We thank the operators of LOAO for their support for our observations. This work includes the data taken with the LOAO 1-m telescope, which is operated by KASI and the SomangNet telescopes.

\pagebreak
\bibliography{ms}

\appendix 
\section{Optical/NIR observations}\label{sec:opt}
\begin{figure*}[h!]
  \centering
  \includegraphics[scale=0.5]{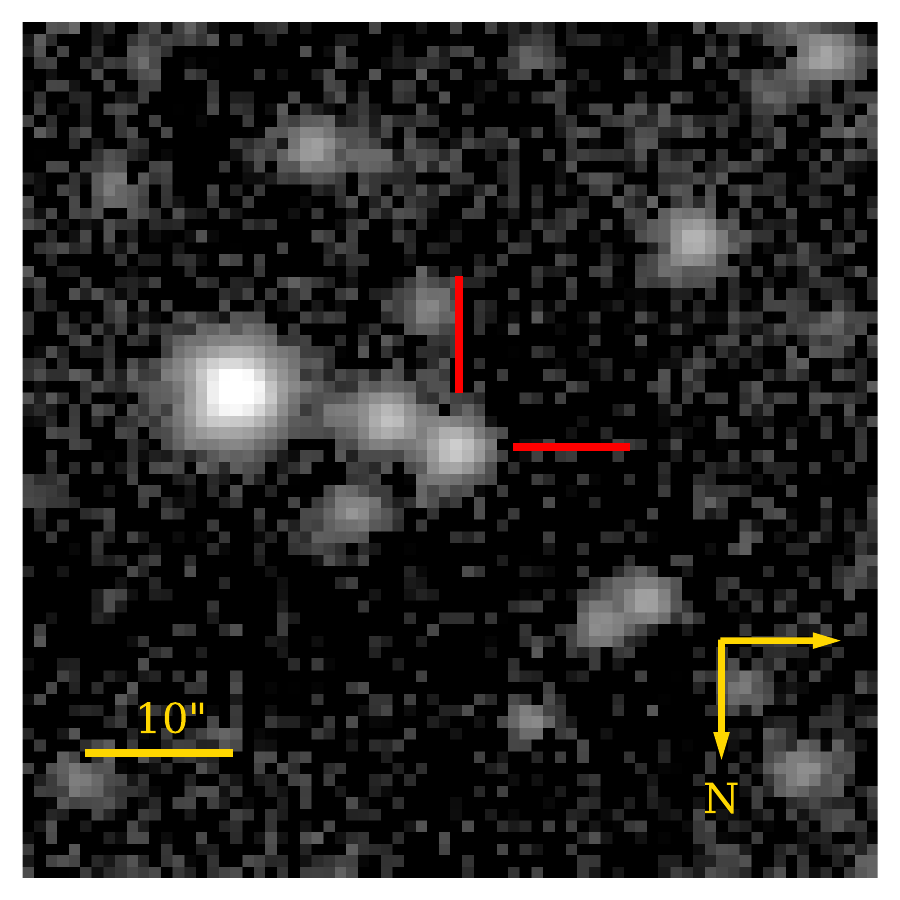}
  \caption{Location of GRB afterglow in I-band at LOAO, 0.63 days (15 hours) after the trigger. Red markers show the location of GRB afterglow.}
  \label{fig:phot}
\end{figure*}

This section provides a summary of the ground-based optical/NIR observations of the GRB afterglow. Table~\ref{tab:phot} summarizes the photometric results. Figure~\ref{fig:phot} shows the location of the afterglow as observed in the I-band of LOAO. All magnitudes are in the AB magnitude system, and the NIR data from BOAO ($J\&Ks$) are converted to an AB magnitude system after calibrating the photometry with the 2MASS catalog point sources \citep{2007AJ....133..734B} by applying offset values $+0.91$ ($+1.85$) for J (Ks)-bands.

\begin{table*}
	\centering
	\begin{tabular}{c c | c c c || c  c | c c c}
		\hline\hline
	Facility & Filter & $t-t_0$ & Mag & Uncertainty & Facility & Filter & $t-t_0$ & Mag & Uncertainty \\
			 &        & [d]         & [AB] & [$5 \sigma$] &    &        & [d]         & [AB] & [$5 \sigma$]\\\hline\hline

	LOAO & B & 0.63 & 20.44 & 0.48 & SAO & V & 2.90 & $\rm >$20.72 & \\
	LOAO & V & 0.63 & 18.70 & 0.12 & SAO & R & 2.94 & $\rm >$20.57 & \\
	LOAO & R & 0.63 & 17.55 & 0.05 & SAO & I & 2.99 & 19.19 & 0.14 \\
	LOAO & I & 0.63 & 16.45 & 0.05 & RASA36 & r & 3.47 & $\rm >$18.86 & \\
	LOAO & z & 0.63 & 16.01 & 0.04 & LOAO & z & 3.59 & 18.69 & 0.14 \\
	LOAO & Y & 0.63 & 15.60 & 0.05 & LOAO & Y & 3.59 & 18.51 & 0.31 \\
	LSGT & g & 0.83 & $\rm >$18.12 & &LOAO & I & 3.59 & 19.55 & 0.11 \\
	LSGT & r & 0.83 & $\rm >$17.77 & &KCT & i & 4.48 & $\rm >$18.09 & \\
	LSGT & i & 0.84 & $\rm >$16.72 & &RASA36 & r & 4.48 & $\rm >$18.77 & \\
	LSGT & z & 0.84 & $\rm >$15.93 & &LOAO & z & 4.59 & 19.00 & 0.14 \\
	KCT & i & 1.45 & $\rm >$17.77 & &LOAO & Y & 4.59 & 18.60 & 0.23 \\
	RASA36 & r & 1.47 & 19.10 & 0.39 &LOAO & I & 4.59 & 20.17 & 0.14 \\
	LOAO & I & 1.60 & 18.01 & 0.07 &RASA36 & r & 5.48 & $\rm >$18.86 & \\
	LOAO & Y & 1.60 & 17.08 & 0.09 &LSGT & i & 5.86 & $\rm >$18.87 & \\
	LOAO & z & 1.60 & 17.43 & 0.05 &LOAO & z & 7.61 & 19.59 & 0.27 \\
	LOAO & R & 1.60 & 19.00 & 0.07 &LOAO & I & 7.61 & 20.86 & 0.25 \\
	SAO & I & 1.89 & 18.23 & 0.05 &BOAO & Ks & 8.45 & 17.66 & 0.05 \\
	SAO & B & 1.91 & $\rm >$20.11 & &BOAO & J & 8.45 & 18.35 & 0.07 \\
	SAO & V & 1.92 & $\rm >$20.11 & &BOAO & J & 9.45 & 18.84 & 0.10 \\
	SAO & R & 1.93 & 19.33 & 0.10 &BOAO & Ks & 9.45 & 18.00 & 0.06 \\
	KCT & i & 2.46 & $\rm >$18.01 & &BOAO & Ks & 10.45 & 18.22 & 0.11 \\
	RASA36 & r & 2.47 & $\rm >$18.75 & &LOAO & I & 11.56 & $\rm >$20.80 & \\
	LOAO & I & 2.56 & 18.85 & 0.07 &BOAO & Ks & 12.45 & 18.46 & 0.06 \\
	LOAO & z & 2.56 & 18.16 & 0.08 &LOAO & z & 12.56 & 20.48 & 0.51 \\
	LOAO & Y & 2.56 & 17.95 & 0.15 &\\

	\hline
	
	\end{tabular}
	\caption{Ground-based Optical/NIR observations of GRB 221009A between 15 hours and 12 days after the trigger. All magnitudes have not been corrected for MW and host galaxy extinction. Upper limits are indicated for non-detection of the GRB afterglow.}
	\label{tab:phot}
	\end{table*}

\section{Posterior probability distributions}\label{sec:app}

As described in Section~\ref{sec:analysis}, we fit the observed optical, X-ray, and gamma-ray data with Equation~\ref{eq:f}. There are 6 free parameters: $K$, $E_b$, $\Gamma_1$, $\log_{10}(E_c)$, $\log_{10}(A_\nu)$, and $N_{\rm H}$. For the second time interval, we add one more free parameter, $C_x$, to take into account the relative calibration factor between \textit{Swift}-XRT and NuSTAR. We use uniform \textit{prior} for $N_{\rm H}$ ($\times 10^{23}$ atoms/cm$^2$), and $C_{\rm x}$, and log-uniform \textit{priors} for other parameters, $K$ (1/$cm^2/s$/keV), $E_b$ (keV), and $E_c$ (keV): $K \in [10^{-2}, 10]$, $E_b \in [10^{-2}, 10^{2}]$, $E_c \in [10^4, 10^8]$, $N_{\rm H} \in [0.8, 1.8]$, and $C_x \in [0.8, 1.2]$. For $\Gamma_1$, we use uniform \textit{prior} for the second time interval, $\Gamma_1 \in [1.4, 1.9]$. We first perform the fit in the second time interval. Then, in the first time interval, we use the Gaussian \textit{prior} based on the marginal posterior distribution obtained from the second time interval, mean and standard deviation of 1.7 and 0.01, respectively. This is because, in our model (Equation~\ref{eq:f}), there is a strong correlation between $\Gamma_1$ and $E_b$, and the spectral index is not expected to evolve significantly within a short time interval. Therefore, we provide more \textit{prior} information for the first time interval. In the case of $\log_{10}(A_\nu)$, we use a Gaussian prior with mean and standard deviation of -0.82 and 0.41, respectively, based on the previous study by \cite{Li2018}.

\begin{figure*}[b!]
  \centering
  \includegraphics[scale=0.5]{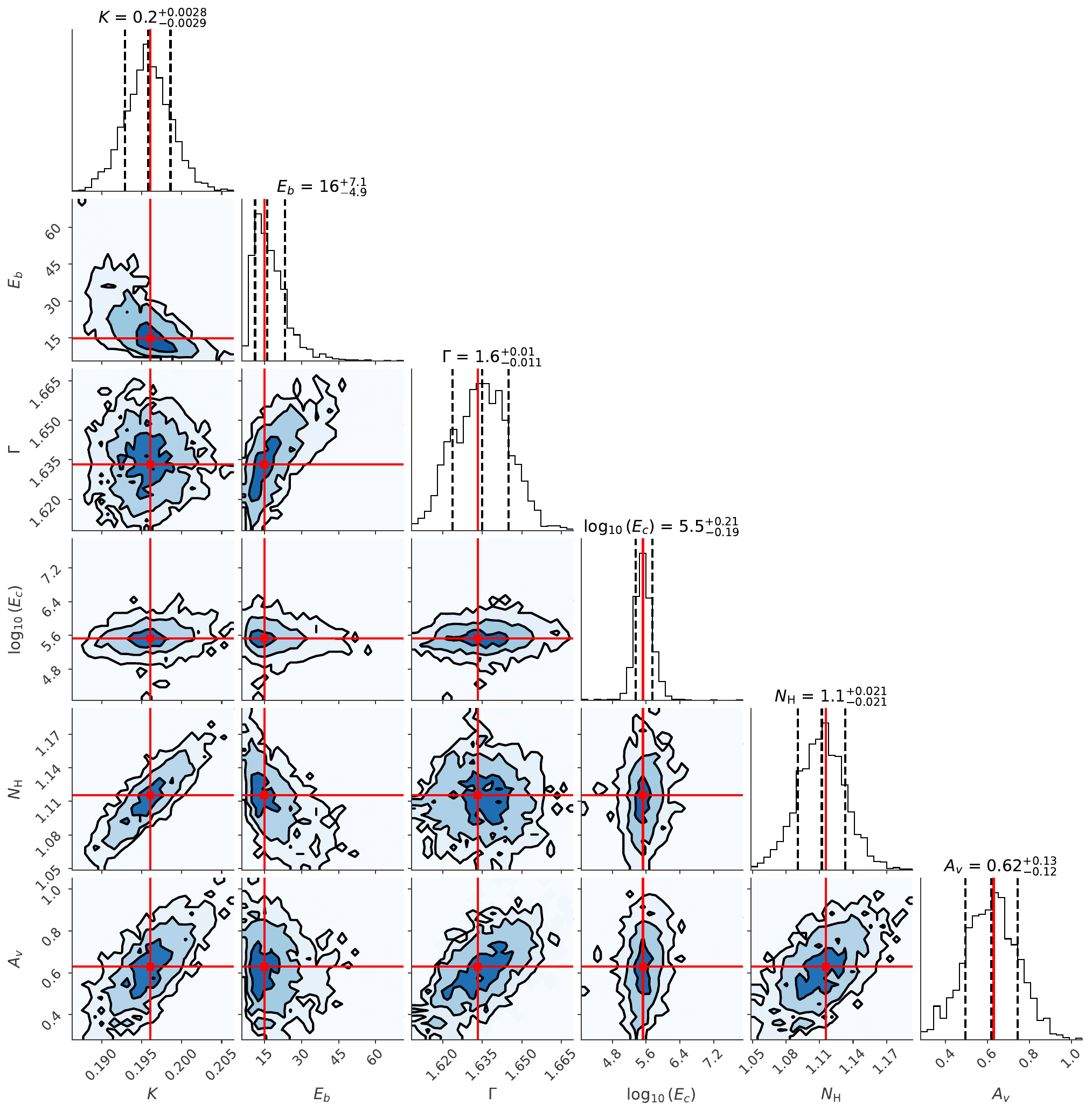}
  \caption{Parameter posterior probability distribution for a time interval from 0.5 to 0.8 days. Dashed black lines correspond to 16, 50, and 84 percentile. Red lines indicate the maximum a posteriori probability point.}
  \label{fig:c1}
\end{figure*}

\begin{figure*}
  \centering
  \includegraphics[scale=0.4]{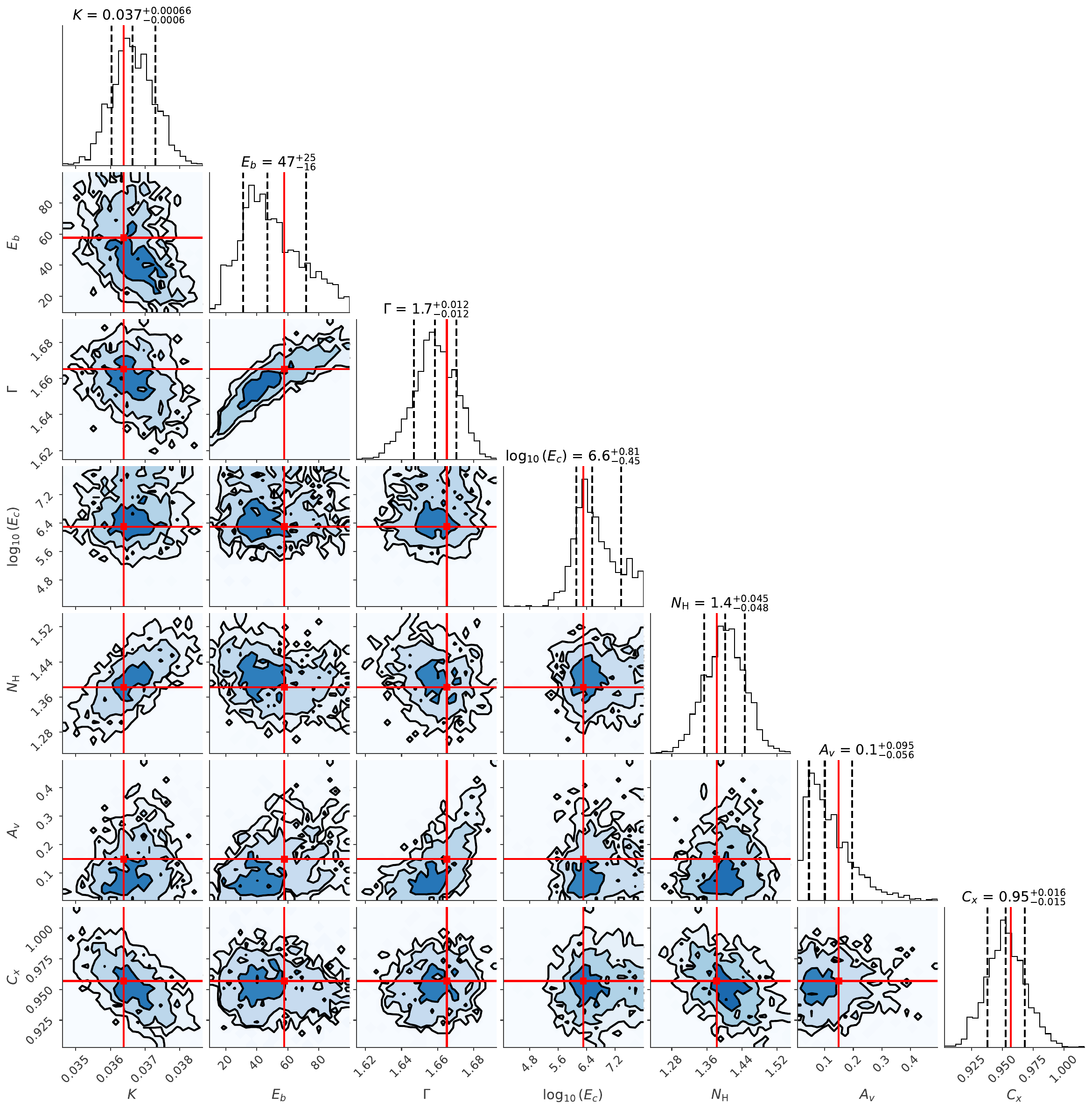}
  \caption{Parameter posterior probability distribution for a time interval from 1.2 to 2.1 days. Notation as in Figure~\ref{fig:c1}.}
  \label{fig:c2}
\end{figure*}

Figure~\ref{fig:c1} and Figure~\ref{fig:c2} present the posterior probability distributions for the first (0.5--0.8 days) and the second time intervals (1.2--2.1 days), respectively. 

\section{Evolution of the maximum synchrotron limit}\label{sec:eb_evol}

\begin{figure*}
  \centering
  \includegraphics[scale=0.65]{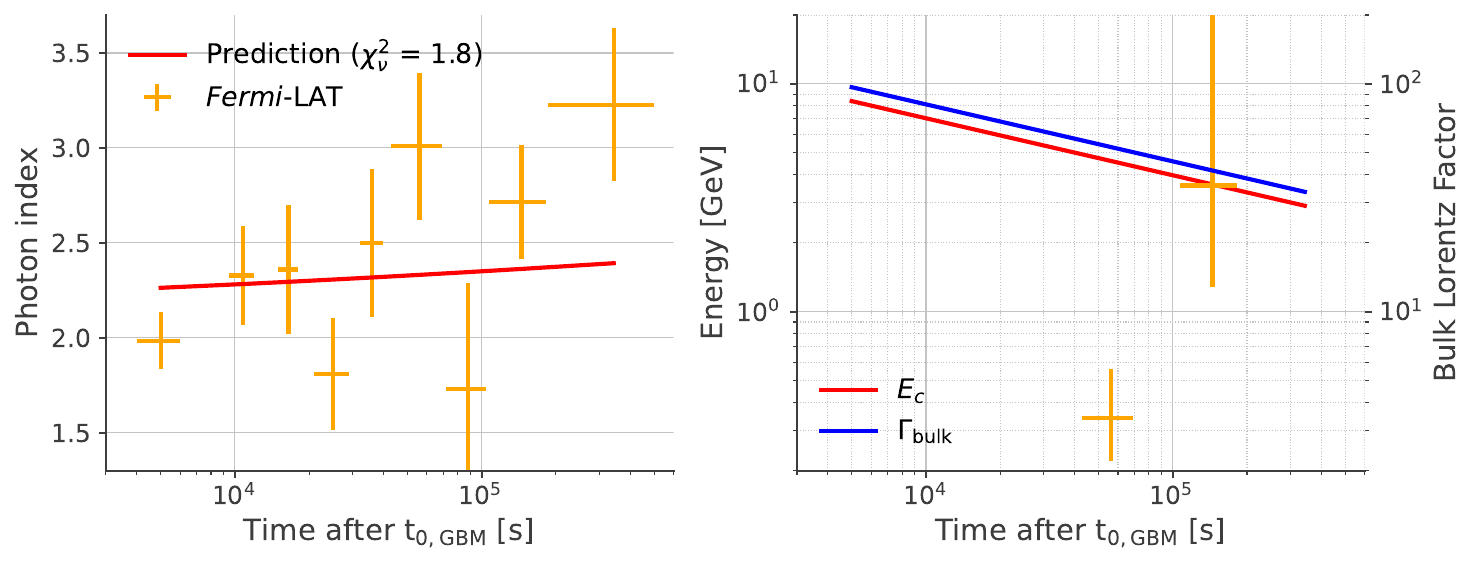}
  \caption{The evolution of the Fermi-LAT photon index and the derived maximum synchrotron limit. The left panel shows the photon index measured in the energy band of 0.1--10 GeV (orange points) and the predicted evolution of the photon index in the maximum synchrotron limit scenario (red solid line). The right panel depicts the cutoff energy (red solid line) and its corresponding bulk Lorentz factor (blue solid line). The orange points in the right panel are from the multi-wavelength modeling (Table~\ref{tab:sed}). }
  \label{fig:Eb}
\end{figure*}

We perform a simulation study to see how the evolution of the maximum synchrotron limit alters the photon index in the \textit{Fermi}-LAT energy band. To do this, we firstly assume that the synchrotron spectrum with its maximum limit can be approximated to Equation~\ref{eq:f}, and $E_c$ corresponds to the maximum synchrotron limit, $E_{\rm max}$. Since the effect of $E_c$ is our interest, we fix the other parameters and synthesize spectra by changing $E_c$ in time, $E_c = E_{c, 0} (t/t_0)^{-1/4}$; $\Gamma_1$ = 1.645, $\Gamma_2$ = 2.145, and $E_b$ = 30 keV. Then, we generate ten flux points in the energy band from 0.1 to 10 GeV and fit the points with a power-law function. As a result, we can see the evolution of the photon index as $E_c$ decays in time. We find the $E_{c, 0}$ value providing the minimum difference between the \textit{Fermi}-LAT data and the simulated evolution. The best-fit value is $E_{c, 0}$ = 8.4 GeV, which is equivalent to $\Gamma_{\rm bulk} \sim 100$, at $t_0$ of $t_{0, \rm GBM} + 5000$ s. The left panel of Figure~\ref{fig:Eb} shows the observed \textit{Fermi}-LAT photon index measured in the energy band from 0.1 to 10 GeV (orange points) and the simulated evolution of the photon index in the same energy band (red solid line). The reduced $\chi^2$ is about 1.8, which indicates a good agreement between the two results. The right panel of Figure~\ref{fig:Eb} exhibits the evolutions of the cutoff energy (red solid line) and its corresponding bulk Lorentz factor (blue solid line) with the results from the multi-wavelength analysis (orange points). The simulation result agrees with the observation result for the second time interval but not for the first time interval. In addition, at the beginning of the external shock deceleration time ($t_{0, \rm GBM} + 120$ s for ISM and  $t_{0, \rm GBM} + 140$ s for wind), as reported by \cite{Lesage2023}, we expect $\Gamma_{\rm bulk}$ to be 245 for ISM and 236 for wind.  

\end{document}